\documentclass[aps,pra,groupedaddress,floatfix,showpacs,twocolumn]{revtex4-1}
\usepackage{amsmath}
\usepackage{graphicx}
\usepackage{bm}
\usepackage{epstopdf}
\usepackage[caption=false]{subfig}
\usepackage{microtype}
\usepackage{placeins}
\usepackage{xcolor}

\begin{document}

\title{How to probe the microscopic onset of irreversibility with ultracold atoms}

\author{R.~B\"urkle$^{1}$}
\email{rbuerkle@rhrk.uni-kl.de}
\author{A.~Vardi$^{2}$}
\author{D.~Cohen$^{3}$}
\author{J.R.~Anglin$^{1}$}

\affiliation{$^{1}$\mbox{State Research Center OPTIMAS and Fachbereich Physik,} \mbox{Technische Universit\"at Kaiserslautern,} \mbox{D-67663 Kaiserslautern, Germany}}
\affiliation{$^{2}$\mbox{Department of Chemistry,} \mbox{Ben-Gurion University of the Negev,} \mbox{Beer-Sheva 84105, Israel}}
\affiliation{$^{3}$\mbox{Department of Physics,} \mbox{Ben-Gurion University of the Negev,} \mbox{Beer-Sheva 84105, Israel}}
\date{\today}

\maketitle
The microscopic onset of irreversibility is finally becoming an experimental subject. Recent experiments on microscopic open and even isolated systems have measured statistical properties associated with entropy production, and hysteresis-like phenomena have been seen in cold atom systems with dissipation (\textit{i.e.} effectively open systems coupled to macroscopic reservoirs). Here we show how experiments on isolated systems of ultracold atoms can show dramatic irreversibility like cooking an egg. In our proposed experiments, a slow forward-and-back parameter sweep will sometimes fail to return the system close to its initial state. This probabilistic hysteresis is due to the same non-adiabatic spreading and ergodic mixing in phase space that explains macroscopic irreversibility, but realized \textit{without} dynamical chaos; moreover this fundamental mechanism quantitatively determines the probability of return to the initial state as a function of tunable parameters in the proposed experiments. Matching the predicted curve of return probability will be a conclusive experimental demonstration of the microscopic onset of irreversibility.

Put a raw egg in a pot on the stove and turn the burner knob; after a while, turn the knob back. The failure of the egg to return to its initial raw state, even though all the control parameters have returned to their initial settings, is a paradigm for macroscopic irreversibility. We may extend the egg-cooking paradigm to microscopic irreversibility by translating it into mechanical terms.
Let a system begin in some preparable initial state and evolve under a Hamiltonian with a time-dependent control parameter which is slowly tuned away from its initial value and then tuned back in exactly the time-reversed manner.  When the control parameter is returned to its initial value, does the system then also return to its initial state? The egg example shows that the answer can often be \emph{No}---in a large and complex dynamical system. How does this form of irreversibility first begin to arise in small systems, which involve only a few degrees of freedom?

\textit{Cooking a microscopic egg.}
This question can be answered in a sufficiently well isolated Bose-Einstein condensate (BEC) in which the interacting bosons can only populate two single-particle states. We assume a Bose-Hubbard form of Hamiltonian with attractive interaction strength $U<0$ and hopping rate $\Omega$: 
\begin{equation}\label{qmH}
\frac{\hat{H}}{\hbar} = -\frac{\Omega}{2}(\hat{a}^{\dagger}_{1}\hat{a}_{2}+\hat{a}^{\dagger}_{2}\hat{a}_{1})+\frac{U\hbar}{2}(\hat{n}_{1}^{2}+\hat{n}_{2}^{2})+\frac{\Delta(t)}{2}(\hat{n}_{1}-\hat{n}_{2})\,
\end{equation}
where $\hat{n}_{j}=\hat{a}_{j}^{\dagger}\hat{a}_{j}$ and $\Delta(t)$ is a tunably time-dependent energy bias between the two states. Hamiltonians of this form have already been realized in cold-atom experiments \cite{Schumm,Obertaler0,Obertaler1,Obertaler2,Trenkwalder}; if dissipation and external noise can be kept negligible over times $T\gg 1/\Omega$ then this kind of system can be used to probe isolated-system irreversibility.
 \begin{figure}
\includegraphics[width=0.5\textwidth]{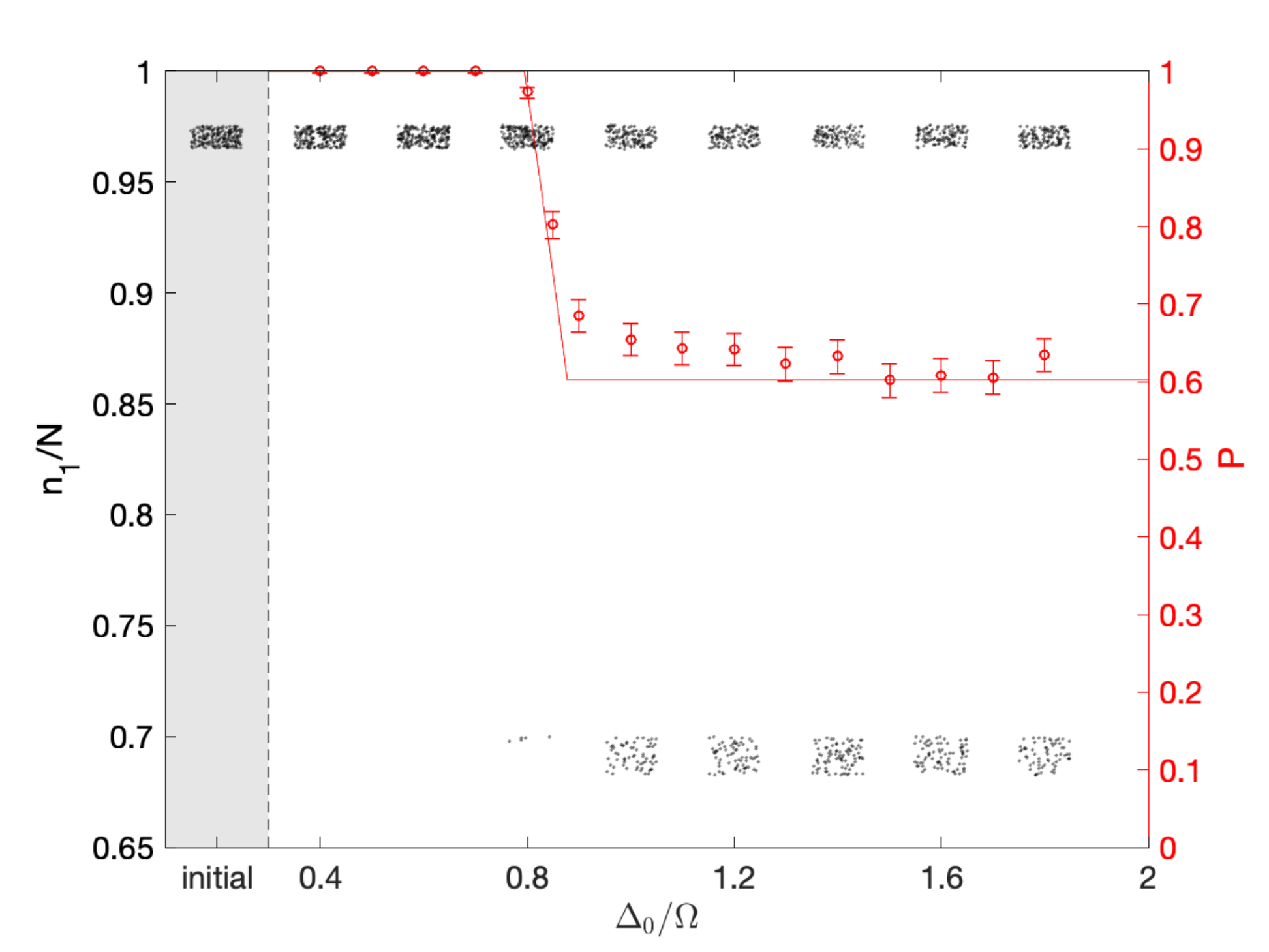}
\caption{Return probability as a function of the sweep extent $\Delta_0$ (open circles, right $y$-axis) as deduced from an initial microcanonical ensemble of 2000 points. Probabilistic irreversibility sets in around $\Delta_0\sim 0.8$. Error bars indicate the expected sampling error. The solid curve is the theoretically predicted return probability $P(\Delta_0,U,\Omega,N)$ for this initial ensemble (see Methods), in the limit $T\to\infty$. Deviations from the theory curve around $\Delta_0\sim 1$ are due to post-adiabatic corrections (see Methods). Black dots show the final particle number $n_1(T)$ (left $y$-axis) for a smaller sample of 200 simulations. All the dots in one column actually have the same $\Delta_0$ but are randomly displaced horizontally to let all the dots be seen. The first column (shaded) shows a typical initial distribution $n_1(-T)$. Probabilistic irreversibility manifests itself as the occurrence of dots with final $n_1(T)$ well outside the initial range.
The parameters for all simulations are $U\hbar N=-3\Omega$, $\Delta_I=-2\Omega$, $-1.68\hbar\Omega N <H(-T)<-1.63\hbar\Omega N$ and constant sweep rate  $(|\Delta_I|+|\Delta_0|)/T = 2 \cdot 10^{-4}\Omega^2$.}
\label{fig:Exp1}
\end{figure}

As the analog for the initial raw egg, let the system be prepared initially in a low-temperature canonical ensemble with large negative $\Delta_{I}$, such that almost all particles are in mode 1. Our microscopic analog for heating and cooling the egg will be to sweep $\Delta(t)$ slowly from an initial value $\Delta_{I}$ at the initial time $t=-T$ to some $\Delta_{0}$ at $t=0$, and then reverse the sweep so that $\Delta(t)$ is again $\Delta_{I}$ at the final time $t=+T$:
\begin{equation}
\Delta(t)= \Delta_{I}\frac{|t|}{T} + \Delta_{0}\Big(1-\frac{|t|}{T}\Big)\;.\label{eq:sweep}
\end{equation} 
Our proposal is then to perform a series of identical experiments with this kind of forward-and-back sweep, similar to the procedures already used in \cite{Trenkwalder}, only without dissipative relaxation to local energy minima. In each experiment the number $n_{1}$ of atoms in mode 1 is measured at the final time $T$ and recorded, so that we obtain a distribution of final values $n_{1}$ for that sweep extent $\Delta_{0}$. The whole series of experiments is then repeated for a set of different values of $\Delta_0$, with no other changes in the procedure. 

With attainably large numbers of particles $N$ a semiclassical theory of evolving a cloud of phase space points under classical equations of motion should be accurate for such a set of experiments, with quantum corrections potentially observable at smaller $N$ \cite{Trimborn, Polkovnikov, Blakie}. We therefore use this approximation to compute the simulated observations shown in Fig.~\ref{fig:Exp1}. Once the semiclassical approximation is established, we deal with a system with only a single degree of freedom (see the Hamiltonian below). In comparison to the traditional thermodynamic limit of very many classical degrees of freedom, as in an egg or a gas under a piston, this is the ultimate microscopic limit.

Apart from the left-most column of points in Fig.~\ref{fig:Exp1}, each point represents the final $n_{1}(T)$ measured in one experimental run. Each separate column of points represents the subset of experiments performed with a different $\Delta_{0}$; the vertical position of each dot shows $n_{1}(T)$, while the horizontal spread of points within each column is simply a random offset used to make the points visible. The left-most column of points shows the distribution $n_{1}(-T)$ in the initial ensemble, which is the same for all runs. To simplify our discussions below, the initial thermal ensemble has been idealized as micro-canonical with a finite energy width; real finite temperature will spread the point distributions vertically.

Up to a certain threshold sweep extent $\Delta_{0}$, the final distribution of $n_{1}$ remains indistinguishable from the initial distribution in every experimental run. Above this threshold $\Delta_{0}$, however, some runs will end with $n_{1}(T)$ far below the initial range. As the sweep extent $\Delta_{0}$ is raised further, the proportion of these anomalous runs rises, until a plateau is reached. Thus for all values of experimental parameters there is a clearly measurable probability $P(\Delta_{0},U,\Omega,N)$ that the final state of the system will be the same as the initial state after the slow forward-and-back sweep. This return probability $P$ is one up to a threshold $\Delta_{0}$, then falls smoothly to a lower plateau. 

This phenomenon is the microscopic onset of irreversibility. If the initial distribution of $n_1(-T)$ corresponds to the raw state of an egg, then the anomalous final state with $n_1(T)$ well below the initial range represents the cooked egg. The continuous onset of irreversibility occurs not through the final state gradually becoming more distinct from the initial state until a macroscopic difference like that between raw and cooked is attained, but rather through an anomalous final state, which even in the microscopic system is quite distinct from the initial state, becoming continuously more probable. To support this interpretation it suffices to look at the semiclassical theory that describes the experiments.

\textit{Phase space picture}. A mean field approximation of $\hat{H}$ \cite{Smerzi} in convenient canonical variables $(q,p)$ (see Methods) is
\begin{equation}
\begin{split}
H=&-\Omega\sqrt{p_{0}^{2}-p^{2}}\cos(q) + U\left(p_{0}^{2}-p^{2}\right)\sin^{2}(q)\\
&+\Delta(t)\sqrt{p_{0}^{2}-p^{2}}\sin(q)\;.
\label{eq:dimerH}
\end{split}
\end{equation}
With the protocol (\ref{eq:sweep}) for $\Delta(t)$, our dynamics under $H$ has Loschmidt time-reflection symmetry about the instant $t=0$: for every solution $q(t),p(t)$ to the equations of motion, $q(-t),-p(-t)$ is also a solution. Individual solutions are in general \textit{not} their own images under time-reflection, however; even in the quasi-static limit where $T\to\infty$ the solutions may have significantly different initial and final energies, because for $U\hbar N/\Omega<-1$ and certain ranges of $\Delta$, $H$ can have an unstable fixed point and a separatrix (see Methods) where the adiabatic approximation breaks down locally. It is important to note that this breakdown of adiabaticity is \emph{unavoidable}: it persists even in the quasi-static limit of infinitely slow sweep rate, because the evolution of the system becomes arbitrarily slow in the vicinity of the unstable fixed point. Thus although the reason for irreversibility of our evolution is ultimately the usual one of adiabaticity breakdown, we have the unusual feature that reversibility cannot be recovered by simply making the external parameter change slower.

\begin{figure}
\includegraphics[width=.5\textwidth]{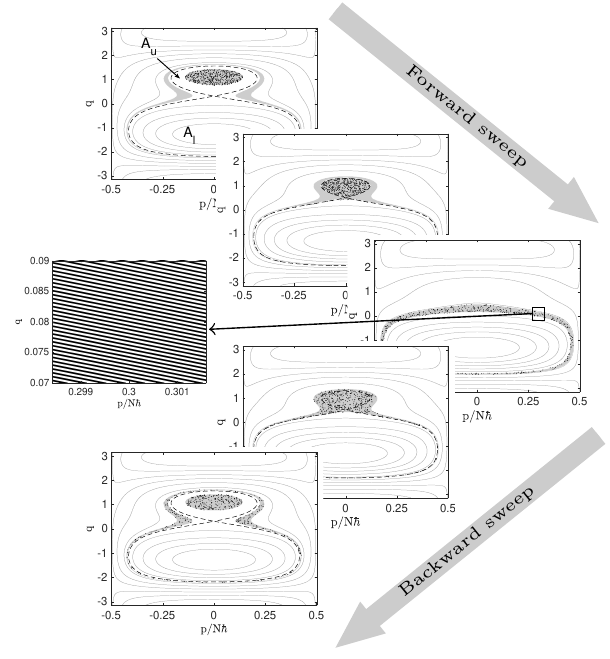}
\caption{Evolution of 500 initial points (black dots) in $(q,p)$ phase space under (\ref{eq:dimerH},\ref{eq:sweep}) with $U\hbar N=-3 \Omega$, $\Delta_I=0.6 \Omega$, $\Delta_0=1.2 \Omega$ and $T=3000\Omega^{-1}$, shown at the times $(-T,-T/2,0,T/2,+T)$. The gray shaded region is the energetic envelope of all the points at time $0$, evolved backward in time to $-T/2$ and $-T$ as well as forward to $+T/2$ and $+T$. Since the shaded region at time $0$ has reflection symmetry in $p$, its evolution has time reversal symmetry, and so it is the same at times $\pm T/2$ and $\pm T$. The dashed line indicates the separatrix; the areas of its $\Delta$-dependent upper and lower lobes are denoted $A_{u,l}(\Delta)$. The left inset panel is a zoom of a small region inside the indicated box in the energy shell at $t=0$, and shows what a \textit{continuous} ensemble that uniformly filled the inner energy shell at $t=-T$ would look like at $t=0$. Within the small zoomed region the stripes are indistinguishable from instantaneous energy contours, but around the energy shell the stripe energies slowly drift: in fact the many stripes are a single continuous swirl that wraps around the shell many times. The initial microcanonical ensemble is thus spread uniformly through the larger, merged energy shell at $t=0$, yet without violating Liouville's theorem. The finite set of black points samples this effectively ergodizing ensemble; the fraction of points that return to the inner shell can be computed in terms of phase space areas via the Kruskal-Neishtadt-Henrard theorem.}
\label{fig:phase_space}
\end{figure}

The effect of the separatrix is illustrated in Fig.~\ref{fig:phase_space}, which shows five instants in the evolution of a sample of phase space points, only some of which return close to their initial state after the slow forward-and-back sweep of $\Delta$. Fig.~\ref{fig:phase_space} also shows the evolution of the energetically allowed phase space shells within which the system points adiabatically orbit. The initial ensemble lies entirely within the upper separatrix lobe denoted $A_u$. As $\Delta$ rises, the separatrix shrinks into the ensemble, which spills non-adiabatically into the lower lobe $A_l$. As this happens, the energy shell within which all points were initially distributed merges with a higher-energy shell that was initially unoccupied. Subsequent adiabatic orbiting distributes the system points throughout the combined shell; even without chaos, the many orbits performed during the slow $\Delta$ sweep effectively swirls the points uniformly throughout the whole energetically available region \cite{Jarzynski,Tiesinga} (left inset). The fact that this effective ergodization occurs with only a separatrix instead of full chaos is surprising but simple (see our Methods), and it makes possible these simple experiments.

In the backwards sweep the single adiabatic shell splits back into disjoint inner and outer shells, in the exact time-reverse of the forward sweep. The distributed points do \textit{not} all find their way back into the inner shell from which they started, however. Instead, because the initial points are effectively randomly distributed throughout the larger shell, the fraction of a thin initial energy shell which finally returns to that shell is given by the ratio of the phase space area of the initial shell to the larger area which the ensemble quasi-ergodically fills at $t=0$. To generalize beyond thin initial energy shells to a general equilibrium phase space distribution function $f(E_i)$ with negligible support outside the initial separatrix, we can simply integrate over initial energy $E_i$. In our Methods we use the Kruskal-Neishtadt-Henrard theorem \cite{Kruskal,Neishtadt,Henrard,LC} to derive the return probability
\begin{equation}\label{eq:Kruskal}
P=1-\int\! \mathrm dE_i\frac{\mathrm dA_i}{\mathrm dE_i}\,f(E_i)  \theta\big(E_i-\bar{E}(\Delta_0)\big)\,\frac{A_l'(E_i)+A_i'(E_i)}{A_l'(E_i)}\;,
\end{equation}
where $A_i(E_i)$ is the area enclosed by the initial energy contour $H(q,p,-T)=E_i$. For each $E_i$, $A_l(E_i)$ is the area that the lower lobe of the separatrix will have at the time when the upper lobe has area $A_u(\Delta)=A_i(E_i)$; note that this implies $A_u'(E_i)<0$. The minimum energy $\bar{E}(\Delta_0)$ is the lowest initial energy contour which will meet the separatrix before the sweep reverses at $t=0$. This formula defines the curve in Fig.~1, for the finite-width microcanonical ensemble specified in the caption.

Our entire scenario is thus a surprisingly simple realization within a two-dimensional phase space of the same dynamical mechanism that makes Joule expansion of a classical gas irreversible. An episode of spontaneous non-adiabatic evolution, during which an initial ensemble expands \textit{into} a larger phase space volume, is followed by effective ergodization which finely mixes the ensemble \textit{throughout} that larger volume, without breaking Liouville's theorem. In the microscopic case the non-adiabatic episode is the separatrix crossing, where adiabaticity breaks down due to intrinsic dynamical instability in the system even though external parameter change remains slow. The subsequent ergodic-like mixing within each energy shell occurs without chaos as the ensemble swirls around the larger shell many times \cite{Jarzynski,Tiesinga}. In the microscopic system the larger space into which the ensemble is mixed is larger only by a modest factor. In the higher-dimensional phase space of a macroscopic system, however, the basic mechanism of expansion and ergodization can easily account for macroscopic irreversibility by bringing the return probability P to near zero. In the latter case chaos is generically involved as discussed in \cite{trimer}.

\textit{Outlook.} The microscopic onset of irreversibility can be observed unambiguously with ultracold atoms as a dramatic phenomenon with final `raw' and `cooked' states that differ by particle numbers large enough to be distinguished without precise atom-counting. Detailed agreement between measured and predicted $P(\Delta_0,U,\Omega,N,T)$ will also confirm that it is really the target phenomenon of microscopic irreversibility that is being observed.

Once this is confirmed, the experimental foothold in the frontier of microscopic irreversibility can be expanded in many directions. Reducing atom number $N$ will strengthen quantum effects \cite{quantum_effects}. The quantum version of our scenario presents a series of avoided level crossings, and the transition probabilities can be calculated by the Landau-Zener formula. In contrast to the classical case discussed here the quantum evolution always becomes reversible in the quasi-static limit. The energy gaps at the avoided crossings are exponentially small in $N$, however, so that already for $N$ of order 10 the sweep time $T$ has to be extremely many times $\Omega^{-1}$ to yield a high return probability. For  realistic sweep times therefore the quantum evolution also becomes irreversible, and the return probability can be compared to the classical values obtained here. Details will be published elsewhere. Letting more than two modes be populated can introduce dynamical chaos \cite{FPE, trimer}. And protocols more complex than the symmetric forward-and-back sweep can seek to test when the microscopic `egg' not only fails to return to its raw state adiabatically but becomes impossible to `uncook' by any available means. The simple case we have shown will be just the beginning.
\vspace{1em}

\textbf{Methods}

\textit{Mean field approximation of (\ref{qmH}):} Instead of using $\hat a_i$ operators we can express the Hamiltonian operator (\ref{qmH}) using the Schwinger angular momentum representation
\begin{equation}
\begin{split}
\hat L_x&=\frac{\hbar}{2} \left( \hat a_1^{\dagger} \hat a_2+ \hat a_2^{\dagger} \hat a_1\right)\\
\hat L_y&=\frac{\hbar}{2}\left(\hat n_1-\hat n_2 \right)\\
\hat L_z&=\frac{i \hbar}{2}\left(\hat a_1^{\dagger} \hat a_2- \hat a_2^{\dagger} \hat a_1 \right),
\end{split}
\end{equation}
giving
\begin{equation}
\hat H=-\Omega \hat L_x+U\left( \hat L_x^2+2 \hat L_y^2 + \hat L_z^2 -\frac{\hat N\hbar^2}{2}\right)+\Delta \hat L_y,
\end{equation}
with $\hat N=\hat n_1 + \hat n_2$. The mean field approximation corresponds to replacing the Hermitian $\hat L_i$ by real variables $L_i$ within each subspace of $\hat N$ eigenvalue $N$. We can then introduce canonical coordinates $(q,p)$ via $L_z=p$ and $L_x+i L_y=\sqrt{p_0^2-p^2}e^{iq}$, where $p_0=N\hbar/2$. In these coordinates the mean field Hamiltonian is (\ref{eq:dimerH}) after dropping an inconsequential constant.

\textit{Unstable fixed point and separatrix:} When an external parameter in a Hamiltonian is slowly varied, the adiabatic theorem ensures the conservation of the action (enclosed phase space area) of each orbit. This means that orbits are smoothly deformed, but the evolution with the time dependence of the external parameter inverted would lead to exactly the same deformation in reverse. The occurrence of irreversibility in our system is due to a localized failure of the adiabatic theorem. This happens because our system can have an unstable fixed point, in whose vicinity the orbital period diverges and the condition for the validity of the adiabatic theorem can never be met. The condition for a fixed point $(q_c,p_c)$ is $\dot q=\partial H/\partial p=0$ and $\dot p=-\partial H/\partial q=0$, which leads in our case to $p_c=0$ and
\begin{equation}
\Omega p_0 \sin(q_c)+2Up_0^2 \sin(q_c) \cos(q_c) +\Delta p_0 \cos(q_c)=0.
\end{equation}
This equation can be solved analytically, but the solutions are lengthy expressions. For any given values of $\Omega, U, N, \Delta$ the equation is also easily solved numerically. It turns out that there always exist at least two dynamically stable solutions (the energy maximum and minimum). For $U\hbar N/\Omega<-1$ and certain ranges of $\Delta$ there exist two more solutions, corresponding to an additional energy minimum and an unstable fixed point \cite{Korsch,DLZ,Niu,Niu2}. The separatrix is then the figure-eight-shaped energy contour passing through the unstable fixed point, as shown in Fig.~\ref{fig:phase_space}. We refer to the two regions that are bounded by the separatrix figure-eight as the upper and lower lobes (see Fig.~\ref{fig:phase_space}); they meet at the unstable fixed point but are otherwise disjoint. In all our $\Delta$ sweeps, the upper lobe may be said to exist from the beginning, with the lower lobe appearing and growing while the upper lobe shrinks. If the sweep extends far enough, the upper lobe eventually disappears. 

The location of the fixed point depends on the bias $\Delta$, and so also does the separatrix energy $E_c=H(q_c,0)$.
The areas of the upper and lower separatrix lobes,
\begin{equation} 
A_u=2\int_{q_c}^{q_{max}}\mathrm d q\; p \left(H=E_c,q \right)
\end{equation}
and 
\begin{equation}
A_l=2 \int_{q_{min}}^{q_{c}}\mathrm d q\; p\left(H=E_c,q\right),
\end{equation}
where $q_{min}$ and $q_{max}$ are the roots in $q$ of $H(q,0)=E_c$, are thus also both functions of $\Delta$.

\textit{Kruskal return probability:}
The Kruskal-Neishtadt-Henrard theorem \cite{Kruskal,Neishtadt,Henrard,LC} deduces the fraction of a thin adiabatic energy shell that will exit a shrinking separatrix lobe into one of two growing regions, from the facts that adiabaticity only fails near the separatrix, and that Liouville's theorem remains valid everywhere. During our forward sweeps the upper lobe is shrinking, and so is the region outside both separatrix lobes, so the theorem prescribes that the orbits being squeezed out of the upper lobe all go into the lower lobe. 

During the reverse sweep, conversely, the lower lobe shrinks while both upper lobe and outer regions are growing, so some orbits migrate into both growing regions of phase space. The phase space area which is squeezed out of the shrinking lower lobe, in a time interval over which $\Delta\to\Delta-\delta$, is $|A_l'(\Delta)|\delta$; meanwhile the upper lobe gains area $|A'_u(\Delta)|\delta$. The share of migrating orbits gained by the upper lobe in this interval is thus $|A'_u|/|A'_l|$.

By our Loschmidt symmetry around $t=0$, the areas $A_{u,l}$ of both lobes when our thin energy shell spills back out of the lower lobe on the return sweep are the same as they were when the shell spilled into the lower lobe on the forward sweep. And since the area initially enclosed by an adiabatic orbit is conserved until the shrinking separatrix lobe shrinks down to meet it, the value of $\Delta$ at which an initial thin energy shell around initial energy $E_i$ will spill through the shrinking separatrix is the $\Delta_c(E_i)$ found by solving $A_u(\Delta_c) = A_i(E_i)$ for $\Delta_c$. This solution then defines both the lobe areas at separatrix crossing $A_{u,l}(\Delta_c(E_i))$ as functions of initial energy $E_i$. 

The ratio of rates of change of lobe areas in the Kruskal-Neishtadt-Henrard formula can then trivially be expressed in terms of derivatives with respect to $E_i$:
\begin{equation}
-\frac{\frac{\mathrm dA_u}{\mathrm d\Delta_c}}{\frac{\mathrm dA_l}{\mathrm d\Delta_c}} = -\frac{\frac{\mathrm dE_i}{\mathrm d\Delta_c}\frac{\mathrm dA_u}{\mathrm dE_i}}{\frac{\mathrm dE_i}{\mathrm d\Delta_c}\frac{\mathrm dA_l}{\mathrm dE_i}}\;,
\end{equation}
yielding the Kruskal-Neishtadt-Henrard return fraction for a thin initial energy shell around $E_i$ as
\begin{equation}
P(E_i) = -\frac{\frac{\mathrm dA_u}{\mathrm dE_i}}{\frac{\mathrm dA_l}{\mathrm dE_i}}
\end{equation}
for all $E_i$ that will actually encounter the separatrix before the sweep reverses at $t=0$. Initial energies below the $\bar{E}(\Delta_0)$ which solves $\Delta_c(\bar{E})=\Delta_0$ will instead have return probability $P=1$, since they never cross the separatrix. Integrating over thin energy shells with the phase space measure $\mathrm dA_i = \mathrm dE_i\,\mathrm dA_i/\mathrm dE_i$ and the ensemble density $f(E_i)$ then yields (\ref{eq:Kruskal}). 
\begin{figure}
\includegraphics[width=0.5\textwidth]{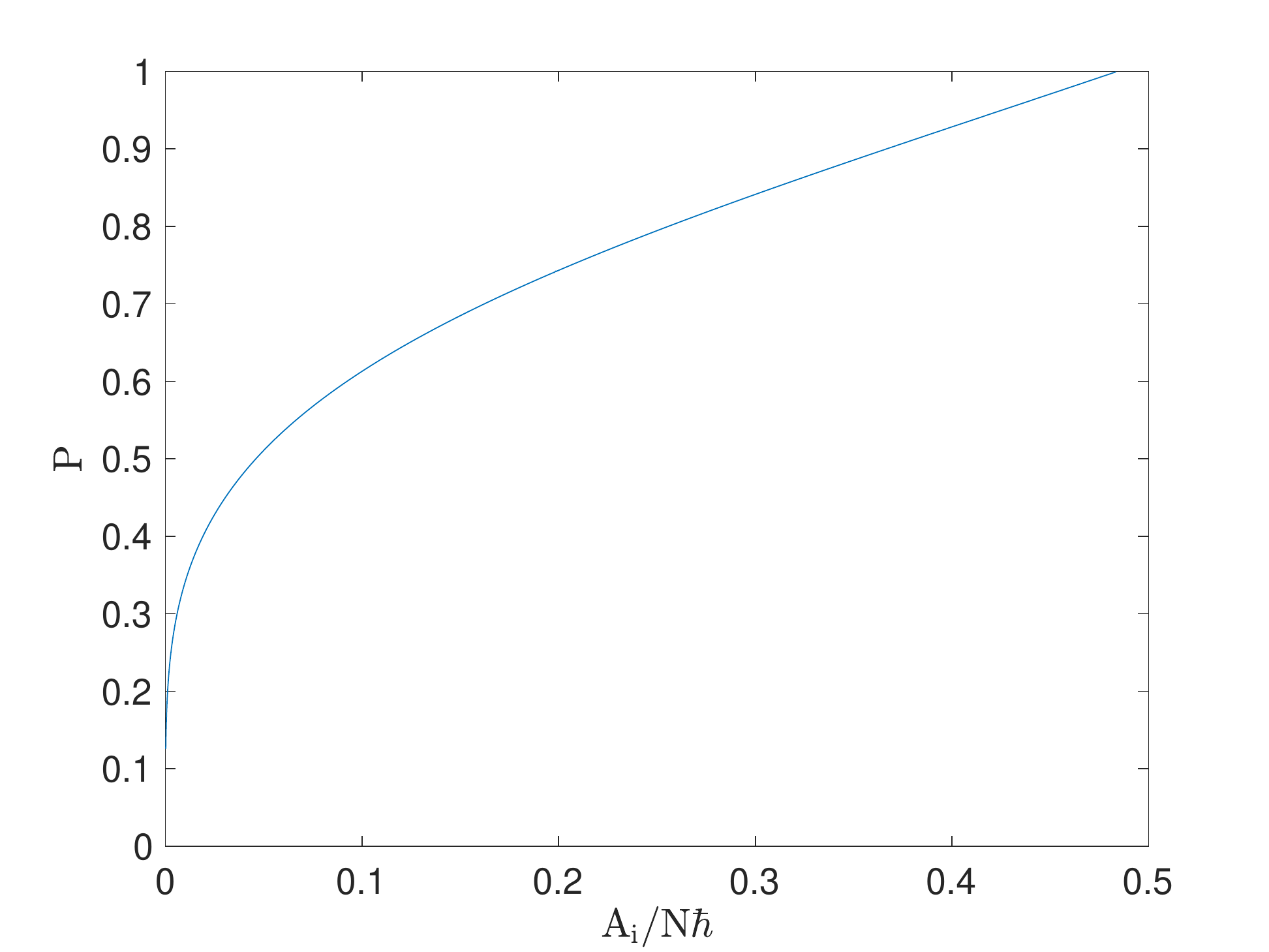}
\caption{Kruskal-Neishtadt-Henrard return probability $P$ for energy contours with initially enclosed area $A_i$ in the upper lobe for $U\hbar N=-3 \Omega$. The return probability can be much lower than in Fig.~\ref{fig:Exp1} if $A_i$ is small enough.}
\label{fig:kruskal_prob}
\end{figure}

If $\Delta_0$ is large enough that a given shell crosses the separatrix, Fig.~3 shows that the return probability $P(E_i)$ can range from near zero to near one. This implies a predictable and observable dependence of return probability on the system's initial temperature.
Note that the nearly straight sloping region of the  $P(\Delta_0)$ curve in Fig.~1 may become less straight for ensembles that are broader in energy. Our simulated data points fit the infinite-$T$ curve less closely for smaller $\Delta_0$ because the system spends more time close to the unstable fixed point in these runs than it does for larger $\Delta_0$. This weakens the adiabatic approximation; neo-adiabatic corrections \cite{Timofeev,Tennyson,Hannay,Cary,Elskens} to provide a more accurate finite-$T$ prediction in this regime may be pursued in future work.

\textit{Swirling vs. coarse graining:}
At $t=0$, just as the sweep is reversing, our ensemble effectively fills the merged energy shell, in the sense that it is swirled throughout the energy shell very finely, as shown in the left inset panel of Fig.~2. A traditional interpretation is that \emph{coarse-graining} makes this finely swirled ensemble equivalent to an ergodic distribution; irreversibility is then explained by asserting that information in the initial ensemble which has been carried into fine-grained features is effectively lost forever. This explanation may be appealing because fine structures can certainly be lost to human perception, but our proposed experimental scenario provides an explicit counter-example to the hypothesis that simple system dynamics cannot reassemble simple shapes out of fine-grained swirls. The reflection in $p$ of our finely swirled ensemble at $t=0$ is exactly as finely swirled as the unreflected ensemble, but our time reversal symmetry implies that the simple separatrix crossing dynamics of our system will reassemble that finely swirled reflected ensemble back into the coarse microcanonical shell that was our initial state. The irreversibility in our scenario is therefore not due to the fact that finely swirled ensembles in general can never dynamically reassemble into coarse ones, because this is not a fact. 

\begin{figure}
\includegraphics[width=0.5\textwidth, trim=8mm 74mm 16mm 80mm,clip]{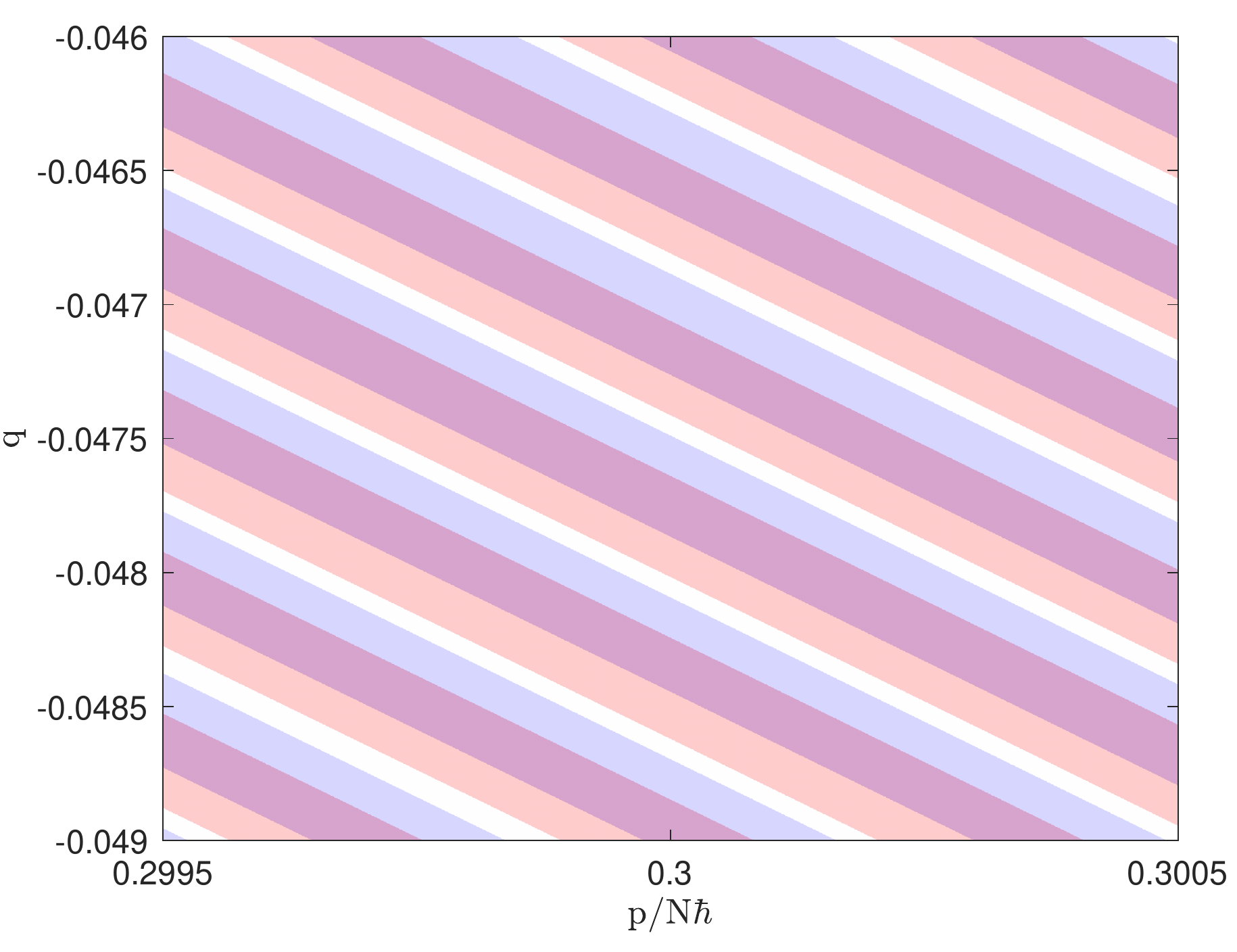}
\caption{Part of the finely swirled ensemble corresponding to Fig.~\ref{fig:Exp1} at $\Delta_0/\Omega=1$ and $t=0$ around $p/N\hbar=0.3$ (blue) and the inverted ($p \rightarrow -p$) ensemble around $p/N\hbar=-0.3$ (orange). The non-unity overlap (violet) is the reason for return probability $P<1$.}
\label{fig:swirling_back}
\end{figure}

Irreversibility occurs instead simply because the finely swirled ensemble $f(q,p,t)$ that we have reached at $t=0$ is not the same as the finely swirled ensemble $f(q,-p,0)$ that would evolve back into our initial state. See Fig.~\ref{fig:swirling_back}. The return fraction for each thin energy shell around $E_i$ can be defined exactly as a classical version of the so-called `Loschmidt echo', as follows. Define the un-normalized initial distribution function $\tilde{f}(q,p,-T;E_i)$ which is simply one within the initial thin shell and zero outside it. Evolve this un-normalized distribution function under the Hamiltonian from $t=-T$ to $t=0$. The exact return fraction for this initial energy shell is then  
\begin{equation}\label{exactP}
P(E_i) \equiv \frac{\int\! \mathrm dq \mathrm dp\,\tilde{f}(q,p,0;E_i)\tilde{f}(q,-p,0;E_i)}{\int\! \mathrm dq \mathrm dp\,\tilde{f}(q,p,0;E_i)}\;.
\end{equation}
The two finely swirled distributions $\tilde{f}(q,\pm p,0;E_i)$ are related simply to each other by $p$-reflection, but this is a global relationship which has nothing in particular to do with the local swirling structure of either distribution. As Fig.~\ref{fig:swirling_back} shows, in any typical small region of phase space the thin stripes of  $\tilde{f}(q,\pm p,0;E_i)$ fail to overlap precisely, so that $P(E_i)<1$. 

Our return formula (\ref{eq:Kruskal}) according to the Kruskal-Neishtadt-Henrard theorem is thus not really based on the naive assumption that the swirled distribution $\tilde{f}(q,p,0;E_i)$ is exactly equivalent to the ergodic distribution filling the $t=0$ energy shell uniformly at coarse-grained density $P(E_i)$. Rather, it is equivalent to assuming that the time-forward and time-reversed distributions $\tilde{f}(q,\pm p,0;E_i)$, both of which are finely swirled, are \textit{ergodic with respect to each other}, in the sense that their overlap with each other in (\ref{exactP}) is the same as the overlap of either with the ergodic distribution. 

\textbf{Acknowledgement}

The authors  acknowledge support from State Research Center OPTIMAS and the Deutsche Forschungsgemeinschaft (DFG) through SFB/TR185 (OSCAR) (JRA and RB) and from the Israel Science Foundation (Grant No. 283/18) (AV and DC).

\textbf{Competing Interests}
The authors declare no competing interests.

\FloatBarrier

\bigskip
\textbf{Contributions}
JRA, AV and DC wrote the main manuscript. RB did the simulations, prepared the figures and wrote the Methods section. All authors reviewed the manuscript.
\end{document}